# Field experiment on a robust hierarchical metropolitan quantum cryptography network

XU FangXing[1], CHEN Wei*[1], WANG Shuang[1], YIN ZhenQiang[1], ZHANG Yang[1], LIU Yun[1], ZHOU Zheng[1], ZHAO YiBo[1], LI HongWei[1,2], LIU Dong[1], HAN ZhengFu†[1] and GUO GuangCan[1]

[1] Key Laboratory of Quantum Information (CAS), University of Science and Technology of China, Hefei 230026, Anhui, China

[2] Electronic Technology Institute, Information Engineering University, Zhengzhou 450004, Henan, China

**A hierarchical metropolitan quantum cryptography network upon the inner-city commercial telecom fiber cables is reported in this paper. The seven-user network contains a four-node backbone net with one node acting as the subnet gateway, a two-user subnet and a single-fiber access link, which is realized by the Faraday-Michelson Interferometer set-ups. The techniques of the quantum router, optical switch and trusted relay are assembled here to guarantee the feasibility and expandability of the quantum cryptography network. Five nodes of the network are located in the government departments and the secure keys generated by the quantum key distribution network are utilized to encrypt the instant video, sound, text messages and confidential files transmitting between these bureaus. The whole implementation including the hierarchical quantum cryptographic communication network links and the corresponding application software shows a big step toward the practical user-oriented network with a high security level.**

Quantum cryptography, quantum key distribution, quantum cryptography network

Supported by Wuhu Government and China Telecommunications Corporation, Wuhu Branch. Supported by National Fundamental Research Program of China (2006CB921900), National Natural Science Foundation of China (60537020, 60621064) and the Innovation Funds of Chinese Academy of Sciences.

doi:

Corresponding author (email: kooky@mail.ustc.edu.cn*; zfhan@ustc.edu.cn†)

In the latest two decades, quantum cryptographic service has gradually been refined due to the rapid improvement of quantum key distribution (QKD) technique in practice [1, 2, 3, 4, 5]. Derived from the fundamental laws of physics, it can offer a highly secure communication between legal users, no matter how the eavesdropper tampers with the system in the open channel [6, 7]. However, traditional point-to-point QKD protocols have their intrinsic limitations on the secure distance and expansibility of users. Quantum secret sharing with GHZ states or W states can give a scheme of the communication for more than two users [8, 9, 10, 11], nevertheless it is more meaningful in quantum communication instead of practical quantum cryptography. For the fast inflation of user number and unforeseen emergent demands of communication service, so far it seems that a robust quantum cryptography network compatible with the classical optical network is a potential solution.

Since the first quantum cryptography idea for passive optical network (PON) is proposed by Townsend et al.

with beam splitters [12, 13], many new topologies for quantum network have been designed and subsequently quantum network construction has been demonstrated in real-built telecom optical network. The first network experiment fielded by BBN with 3 nodes in Boston is based on the active optical switches [14]. Following that, we built a 4-port "star type" quantum network in Beijing benefited from a "quantum router" structure which can realize the passive routing with WDM apparatus [15]. SECOQC also constructs the network with 7 QKD links to connect 5 nodes in Vienna [16]. Recently, Chen et al. applied the decoy state method to connect 3 nodes together by treating the node in the middle as a trusted relay [17].

Here we give a report about our brand-new hierarchical quantum cryptographic network with decoy state method located in Wuhu, China, operated continuously beneath the streets of the city. With the improved networking techniques of the quantum router, optical switch and trusted relay, 7 subsidiaries of the city are connected to form a secure communication network.

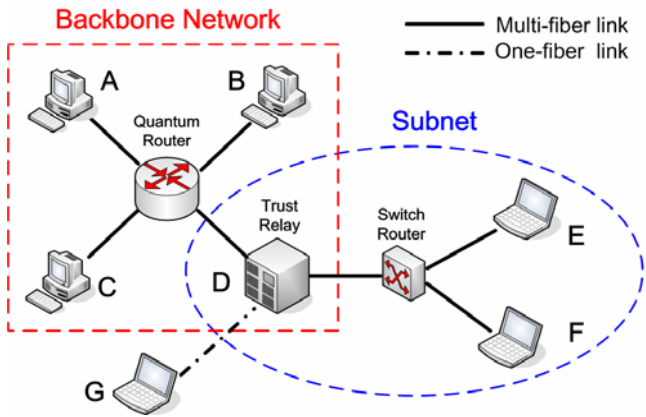

Figure 1 Structure of the hierarchical quantum cryptographic network, which contains techniques of quantum router, trust relay and switch router. A to G stand for seven different terminal nodes, which link to the backbone in the red dash square or the subnet in the blue dash circle individually. The straight line in the figure means an optical link with multi-fiber, while the dot dash line means the optical link with a single fiber.

The diagram in Figure 1 gives a detailed exhibition about the quantum network. In this network, we construct different levels after considering the priority of the bureaus. For four important nodes, a high-level full mesh backbone network is built among them and each of them can also operate as a trust relay to expand the net. Two of the others belong to a subnet, link with an optical switch matrix to one trust relay node shown as Node "D" in the diagram of the backbone network. In addition, we use one single telecom fiber to add the seventh node into the network, for both the classic network connection and QKD distribution, to infer the potential of our QKD network even if the fiber channel is limited.

Compared with the prior network projects, this QKD network implements hierarchical structure with multi-levels for the first time and contains three different existing PON techniques. For nodes with different priorities and demands, we set them in the central backbone network or the subnet, and choose suitable PON technique. All the QKD links are based on the BB84 protocol with decoy state method which can promise the security level for the communication. Meanwhile, QKD software that all nodes run and application programs for encrypting text messages, sound and video are developed as well. So this quantum cryptographic network is not just a prototype but a practical user-oriented one.

## 1. Technical introduction
### 1.1 Basic QKD-link device

Each path has one QKD link with a set of devices between two terminals. Here, we use the Faraday-Michelson Interferometer (FMI) set-up to implement the phase coding QKD [3]. Jeopardized by the polarization dependent loss (PDL) and polarization mode dispersion (PMD) of the single mode communication fibre, most of the QKD implementations have to use a polarization controller (PC) to stabilize the polarization and decrease its influence in laboratory experiments. Nevertheless in a real station node, it is not convenient to plug such a device into the well-compacted system box. Significantly, FMI QKD set-up can take advantage of the Faraday mirrors to constrain the influence of the polarization in the way [7, 18]. A Faraday mirror's Jones matrix can be given as

$$FM = \begin{bmatrix} 0 & -1 \\ -1 & 0 \end{bmatrix} \qquad (1)$$

Subsequently the transmitting Jones matrix in the chanel $T_{all}$ can be expressed as

$$\bm{T}_{all} = \bm{T}^{-1} \cdot \bm{FM} \cdot \bm{T} = e^{i\varphi} \bm{FM} \qquad (2)$$

Here $T$ stands for the unidirectional transmittance that is modeled as consisting of a discrete number of birefrigent elements. The analysis shows that the output polarization is always orthogonal to the input one regardless of any birefringence in the fiber, and guarantees that the FMI QKD setup can auto-compensate the polarizing influence from the channel fluctuation, which has been already proved performing well in the metro telecom fibers combining the active phase compensation technique [3, 15, 19].

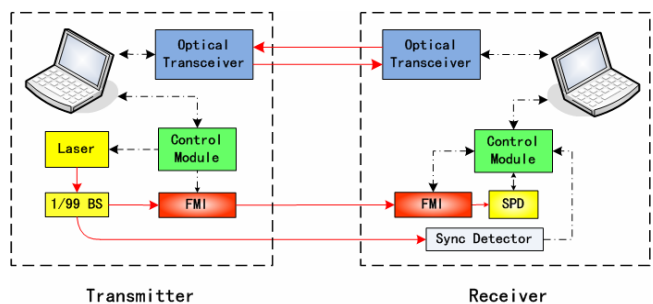

Figure 2 Scheme of the original FMI system. Quantum signal pulses and synchronization pulses are generated by the laser and a 1: 99 beam splitter (1/99 BS), which occupy one fibre each. Adding to two fibres for optical transceivers links, four fibres are required in total shown with red lines. The dash dot lines stand for the electrical signals.

The communication resource needs to be well concerned as well. In the traditional optical network, with the concern of cost, only the important nodes with requirements of high key generation rate and low congestion are distributed enough fibers and channels. A none-essential station maybe has only one or two fibers to connect with the backbone network. Consequently, we have demonstrated two different schemes for these two kinds of nodes respectively. Scheme one shown in Figure 2 is the original FMI system, that quantum signal pulses and synchronizing pulses transmit in two independent fibers with the same wavelength. Additionally, we use the commercial bi-channel optical transceivers to build the classic communication link. By sending the 1550nm strong pulse to the receiver, one couple of them can well communicate with each other after tolerating 30dB channel loss (150 km in the standard communication fibres). Typically this scheme needs four independent fibres. The advantage is that without the crosstalk from other pulse sequence, the QKD process can give a nice performance with lower quantum bit error rate (QBER) and subsequently higher key generation rate. This scheme is suitable to distribute quantum secure key between important nodes with sufficient communication fibers and other resource.

For those low-level access nodes, we use the modified scheme shown in Figure 3. Compared with scheme one, four different pulses sequences (classical transmit pulses, classical receiving pulses, quantum signal pulses and synchronizing pulses) are multiplexed into one fibre, by using a circulator and a wavelength division multiplexing (WDM) element. It will significantly save the communication resource of the optical network. However, the influence of the Raman scattering and direct crosstalk will evidently increase the QBER and restrict the communication ability [20], which can be observed in the field-test data below.

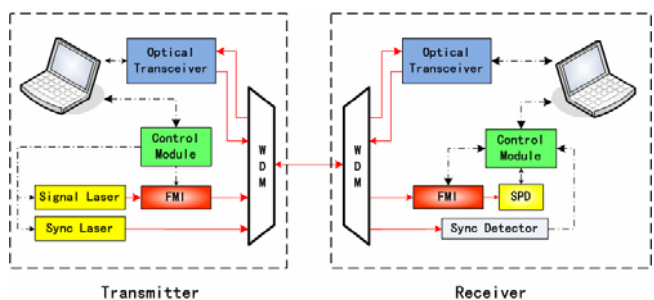

Figure 3 Scheme of the modified FMI system. Quantum signal pulses and synchronization pulses are generated by lasers with distinguishable wavelengths. Each of the users multiplexes and demultiplexes the pulse sequences with a WDM element and transmits the photons in one single fiber between each other.

### 1.2 Building a backbone network

Backbone network as the main infrastructure in the whole network should have enough ability to tie the subnets together. So the network capacity and congestion are taken into consideration with high priority, instead of the high data throughput rate.

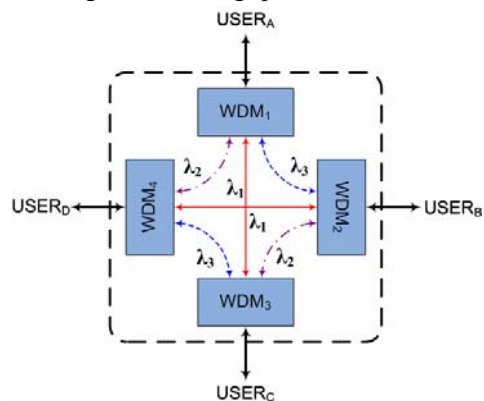

Figure 4 Structure of a four-port quantum router (QR). User A uses photons of different wavelength $\lambda_1$, $\lambda_2$ and $\lambda_3$ to link with users B, C and D. The router can automatically transmit the photons to different outputs for corresponding users. Other users follow the same manipulations to build the QKD links.

Quantum router (QR) with WDM elements in Figure 4 can supply a direct passive routing, which means that each user can distribute the quantum keys with any oth-ers without any manual control. For different individuals, one can use photon pulses with specific wavelength to communicate. That is, in a N-port network with QR in the centre, N(N-1) channels can be established at the same time [21]. Meanwhile, because of the high extinction loss of WDM elements, we can ignore the crosstalk from other channels and actualize all the links simultaneously [15]. The significance of this quantum router structure is that it not only inserts small loss in the channel but also can avoid the receive-resend process that is easy to cause the information delay in the network with the classical trusted relay, which is quite suitable to construct a backbone network with the demand of multiple communication links simultaneously. So far, with the off-the-shelf devices a practical quantum cryptography network containing more than 100 key users can be built, and more users can be added into the subnets flexibly.

### 1.3 Building a subnet

Subnet is an important concept in the construction of a practical network, which comes from the TCP/IP protocol. Hierarchically arranged, subnets can separate the whole network into smaller efficient parts to prevent the information collision in the backbone part. In a subnet, we can compromise on the delay of data transmittance to achieve high transmittance efficiency and cost efficiency.

The subnet part of our quantum cryptographic network in Figure 1 is a typical one served with an optical switch router and a trusted relay. The server in Node D is the border between the backbone net and the subnet and controls the traffic as a trusted relay, while the switch router is controlled by the computer to establish the connection for D and E or for D and F, both in the quantum signal channel and the synchronization channel. The scheme of network with optical switch is shown in Figure 5.

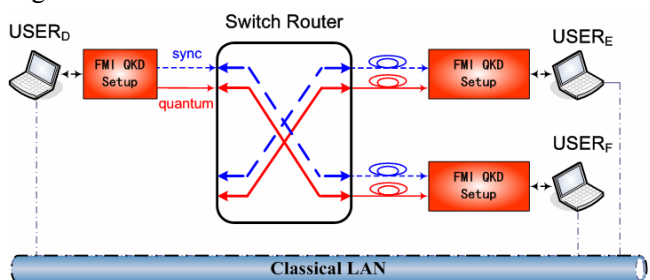

Figure 5 Scheme diagram of a three-user subnet built with an optical switch. The channels for quantum signals and synchronization pulses are switched simultaneously by the optical switch for the QKD link between different users. A classical local area network is built in advance with optical transceivers to link all the users in the subnet.

Optical switch router gives a simple and easy-to-implement solution to the expandability of the network [14, 22]. However, as long as a shared channel can only be used for one QKD connection in such a network structure, we have to keep a close watch on the information congestion and network complexity. That infers that the high-efficiency channel control and key management will become important when lots of optical switches are used in one net [23].

Trusted relay technique is widely used to extend the communication distance [24, 25]. The principle is that all information travelled between the subnet and the backbone net is totally known by the trusted relay. Once the security of these trusted relays is guaranteed, the QKD network is secured. Meanwhile, for a coherent state QKD and decoy state method, trusted relay is quite compatible with the scheme. So far trusted relay has been a suitable technique in a large-scale quantum cryptography network.

### 1.4 Network protocols

When we talk about a quantum cryptographic network, it actually contains two parts in principle. One is the pure optics-connected network to transmit photons and distribute quantum keys. The other is the local area network (LAN) for the classical communication which is necessary for QKD process. This classical network is compatible with an Ethernet or Internet as long as the communication module follows the standard TCP/IP protocols. In this Wuhu quantum network, considering that there is no Ethernet access point in the telecom stations, we use transceivers to build our own LAN, with the schemes sketched above in section 1.1.

As for the QKD process, BB84 phase coding protocol with weak+vacuum decoy state method is implemented for all the links of the network. In the decoy states method, the decoy pulses are designed to sneak into the signal photons randomly. Because it is postulated that Eve cannot distinguish the signal and decoy pulses, the loss and the error rate of signal states can be estimated by the measurement result of the decoy states [26, 27, 28, 29]. Following ref. [30] (also known as GLLP), the final key generation rate is calculated as

$$R \geq q\{-Q_\mu f(E_\mu)H_2(E_\mu) + Q_1^{L,\nu,0}[1 - H_2(e_1^{U,\nu,0})]\}$$

(3)

where $q$ is the efficiency of the FMI scheme, $f(e)=1.2$ is the bidirectional error correction efficiency and $H_2(x) = -xLog_2x-(1-x)Log_2(1-x)$ is the binary entropy function. $Q_\mu$, $E_\mu$ are the gain and error rate of signal states with average photon number $\mu$, while $Q_1$, $e_1$ are the gain and error rate of single photon states which can be estimated as follows [31],

$$Q_1 \geq Q_1^{L,v,0} = \frac{\mu^2 e^{-\mu}}{\mu v - v^2}\left(Q_v e^v - Q_\mu e^\mu \frac{v^2}{\mu^2} - \frac{\mu^2-v^2}{\mu^2}Q_{vac}\right)$$

(4)

$$e_1 \leq e_1^{U,v,0} = \frac{\mu e^{-\mu}}{\mu-v}\frac{E_\mu Q_\mu e^\mu - E_v Q_v e^v}{Q_1^{L,v,0}}$$ (5)

When the statistics fluctuation of the measurement is considered, $Q_v$, $Q_{vac}$ should be replaced by their upper or lower bounds correspondingly shown as follows,

$$Q_x^L = Q_x(1-\frac{10}{\sqrt{N_x Q_x}})$$ (6)

$$Q_x^U = Q_x(1+\frac{10}{\sqrt{N_x Q_x}})$$ (7)

where $x=\mu$, $v$, $vac$, stands for signal state, decoy state or vacuum state, and $N_x$ is the pulse number respectively. In our test, because the data size $N_x$ is large enough to make the difference between the bounds $Q_x^L$ and $Q_x^U$ as small as $0(10^{-2})Q_x$, we can ignore the fluctuation and use the gain in the calculation directly. Meanwhile, even though the extinction ratio of the intensity modulator that we used to switch the pulses among such three different states is 25dB at the most, the so-called vacuum state is not a real vacuum state. So it is more appropriate to estimate the upper bound of $e_1$ tightly with $Q_v$ and $E_v$ in inequality (5) instead of $Q_{vac}$.

With the measurement result achieved in the metro quantum network, we can use the above equations to estimate the key generation rate for the post-processing including CASCADE error correction and privacy amplification with a 2-universal hash function, and distribute a stream of secure keys between terminals which can be utilized in practical applications for legal users.

## 2. Field operation test

The whole quantum cryptographic network is built on the inner-city telecom fiber cables with the distribution in the satellite map of Figure 6. In such a field instead of laboratory environment, the emergent influences of the external vibration and the disturbance from classical optical communications have given a crucial test of the robustness and stability of this QKD network. Nevertheless, our test data clearly shows that the implementation of this quantum cryptographic network is good enough for practical utilization.

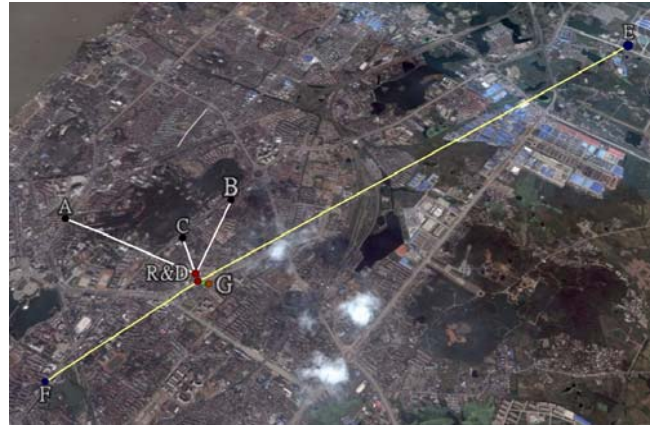

Figure 6 Distribution map of the nodes in the quantum network in Wuhu, Anhui. Node A locates in Bureau of Science and Technology. B stands for the building of Economic Committee. C is in the office of General Labour Union. R stands for the quantum router, which locates in the same telecom station with Node D. The backbone network is composed of these four nodes. E and F belong to the subnet, and are settled in the Bureau of Quality and Technical Supervision and the Civic Commerce Stimulus Bureau respectively. G, in the telecom station as well, is used to show the network compatible with a single-fiber connection.

We have operated the whole implementation continuously to check its performance. All the devices for QKD run with the system frequency of 5 MHz. Single photon detectors, including id200 SPD (idQuantic) and homemade SPD, have similar quantum efficiency of 10% and dark counts of nearly $10^{-6}$. The visibility is 98.67% on average. According to the intrinsic error of the FMI set-up and a single photon detector, the average photon number of signal state and decoy state are unified as $\mu=0.6$ and $v=0.2$ for all links [31], and the ratio among signal state, decoy state and vacuum state is set as 6: 3: 1. The QKD process distributed the sifted key with errors, and subsequently we can distill the final key with a Cascade error correction algorithm and a hash privacy amplification using parameters from the decoy state method.

Table 1 shows the parameters and final key generation rate for all nine links calculated from the full record of the experimental data in Table 2 in Appendix. Both of the fiber length and the insertion loss are measured. Node D and E depart farthest of 10 kilometres, but the link between Node A and C gives the largest loss of 7.18 dB, which also indicates that in a practical fiber network the connection loss is also the main source of the channel loss, except the transmittance loss of the fiber cable with the coefficient as 0.2dB/km. Node A to F is distributed with quantum router or optical switch, using the original FMI set-up with four fibers, and the QBER for signal states is 1.8% ~ 2.4% steadily in a metro scale of 10 km. Both the sifted key generation rate and the final key rate are shown as well. Corresponding to different insertion losses, the final key rate changes from 2.5 Kbps to 0.5 Kbps. The last route from Node D to G is linked by a single fiber of 500 meters with the insertion loss of 1 dB. The QBER has obviously increased to 3.8% because of the crosstalk and Raman scattering in the channel, which causes the final key rate of only 83 bps even though the sifted key is as large as 11.0 Kbps. The result indicates that for a QKD process, both the sifted rate key and QBER are important for the final key gain. The sifted key rate depends on the efficiency and speed of the scheme and the routing technique in a network, while the performance of SPD and the balance of the setup affect the QBER. For a higher key generation rate in the future, we will focus on improving the speed of the whole system and refining the SPD with better behaviour.

| Route | A-R-B | A-R-C | A-R-D | B-R-C | B-R-D | C-R-D | D-S-E | D-S-F | D-G |
|---|---|---|---|---|---|---|---|---|---|
| Wavelength (nm) | 1550 | 1530 | 1510 | 1510 | 1530 | 1550 | 1530 | 1530 | 1510 |
| Distance (km) | 5.0 | 5.6 | 3.5 | 3.6 | 1.5 | 2.1 | 10.0 | 5.0 | 0.5 |
| Attenuation (dB) | 6.28 | 7.18 | 4.42 | 5.13 | 2.39 | 4.37 | 6.23 | 6.14 | 1.0 |
| Sifted Key (Kbps) | 3.38 | 2.56 | 5.32 | 4.36 | 8.25 | 5.42 | 3.15 | 3.27 | 11.0 |
| QBER | 2.19% | 1.87% | 1.96% | 2.15% | 1.93% | 1.8% | 2.34% | 1.89% | 3.8% |
| Final Key (Kbps) | 0.74 | 0.61 | 1.73 | 0.82 | 2.53 | 1.58 | 0.49 | 0.66 | 0.08 |

Table 1, Parameters and the key generation rate of the entire route paths in the quantum network. The row of route stands for the pathway of one link, in which A to G stand for seven terminal nodes, R is the quantum router and S is the optical switch router. The row of Distance is the actual length of the fiber cables which is not proportional to the practical channel attenuation due to the connection loss and other factors.

With the current quantum network system, the key generation rate is insufficient for the one-time pad cryptography utilities, especially when encrypting a video and sound stream with massive data. Consequently we use the Advanced Encryption Standard (AES) algorithm, a classical symmetrical cryptography algorithm to encrypt the plaintext with the fast refreshed key sequence of 128 bits supplied by the QKD process. Based on that, a practical netmeeting is demonstrated for the Bureaus connected in this network. The sound and video stream signals are transmitted after AES encryption to specific subsidiaries through the classical LAN for the instant conference. Secure transmissions for text messages and confidential files can also be actualized with well-developed software that is easy to operate for users.

## 3. Conclusion

In May 2009, we demonstrated the first hierarchical metropolitan quantum cryptography network with seven nodes on the inner-city commercial telecom fibers, realized by two different kinds of FMI set-ups, operated continuously beneath the streets of Wuhu downtown of China. For a high running efficiency, the whole network is divided into two parts with different priorities, a full-mesh backbone net with a quantum router in the center and a subnet based on the technique of the optical switch and the trusted relay, which can well guarantee the feasibility and expandability of the quantum network. Meanwhile, we utilize the secure key distributed by the quantum network in a practical video conference for all the nodes including the transmittance for instant video, sound, text messages and confidential files. The whole implementation with hierarchical QKD network links and well-developed application software clearly shows a big step toward the practical user-oriented quantum network.

## APPENDIX: EXPERIMENT DATA

Table 2, Field test result of the quantum cryptography network. The row of route stands for the pathway for each link. $N_X$, $Q_X$, $E_X$ are the pulse number, gain and QBER for $X$ state, where $X=\mu, v, vac$ stands for the signal state, decoy state and vacuum state respectively. All the data is achieved during a continuous running period of 20 minutes.

| Route | A-R-B | A-R-C | A-R-D | B-R-C | B-R-D | C-R-D | D-S-E | D-S-F | D-G |
|---|---|---|---|---|---|---|---|---|---|
| $N_\mu$ | $1.25\times10^9$ | $1.40\times10^9$ | $1.08\times10^9$ | $1.23\times10^9$ | $1.11\times10^9$ | $1.23\times10^9$ | $1.18\times10^9$ | $1.08\times10^9$ | $1.20\times10^9$ |
| $Q_\mu$ | 0.0069169 | 0.0052029 | 0.0108466 | 0.0085879 | 0.0167524 | 0.0106854 | 0.0063737 | 0.0065981 | 0.022125 |
| $E_\mu$ | 0.0219 | 0.0187 | 0.0196 | 0.0215 | 0.0193 | 0.0180 | 0.0235 | 0.0189 | 0.0380 |
| $N_v$ | $6.27\times10^8$ | $7.00\times10^8$ | $5.40\times10^8$ | $6.14\times10^8$ | $5.53\times10^8$ | $6.14\times10^8$ | $5.90\times10^8$ | $5.41\times10^8$ | $6.02\times10^8$ |
| $Q_v$ | 0.0024508 | 0.0018022 | 0.0040541 | 0.0029451 | 0.0061579 | 0.0038264 | 0.0021723 | 0.0022359 | 0.007593 |
| $E_v$ | 0.0307 | 0.0312 | 0.0331 | 0.0309 | 0.0326 | 0.0330 | 0.0306 | 0.0278 | 0.0425 |
| $N_{vac}$ | $2.09\times10^8$ | $2.33\times10^8$ | $1.80\times10^8$ | $2.05\times10^8$ | $1.84\times10^8$ | $2.05\times10^8$ | $2.05\times10^8$ | $1.80\times10^8$ | $2.01\times10^8$ |
| $Q_{vac}$ | 0.0001074 | 0.00009 | 0.0001233 | 0.0001539 | 0.0002112 | 0.0001586 | 0.0001180 | 0.0001319 | 0.000396 |
| $E_{vac}$ | 0.2016 | 0.1564 | 0.2323 | 0.1485 | 0.1994 | 0.1947 | 0.1575 | 0.1511 | 0.1588 |
| Time | 1283 s | 1426 s | 1103 s | 1214 s | 1125 s | 1213 s | 1196 s | 1094 s | 1212 s |